\documentclass[conference]{IEEEtran}
\usepackage[utf8]{inputenc}
\usepackage{graphicx}
\usepackage{amsmath,amssymb,amsfonts}
\usepackage{multirow}
\usepackage{booktabs}
\usepackage{xcolor}
\usepackage{caption}
\usepackage{subcaption}
\usepackage{cite}
\usepackage{url}
\usepackage{microtype}
\usepackage{algorithm}
\usepackage{algpseudocode}

\title{Rethinking Hybrid Retrieval: When Small Embeddings and LLM Re-ranking Beat Bigger
Models}

\author{
  \IEEEauthorblockN{Arjun Rao\quad Hanieh Alipour\quad Nick Pendar}
  \IEEEauthorblockA{SAP, San Ramon, USA\\
    \{arjun.rao01, hanieh.alipour, nick.pendar\}@sap.com}
}

\begin{document}
\maketitle

\begin{abstract}
This paper presents a comparison of embedding models in tri-modal hybrid retrieval for Retrieval-Augmented Generation (RAG) systems. We investigate the fusion of dense semantic, sparse lexical, and graph-based embeddings, focusing on the performance of the \textbf{MiniLM-v6} and \textbf{BGE-Large} architectures. Contrary to conventional assumptions, our results show that the compact \textbf{MiniLM-v6} outperforms the larger \textbf{BGE-Large} when integrated with \textbf{LLM-based re-ranking} within our tri-modal hybrid framework. Experiments conducted on the SciFact, FIQA, and NFCorpus datasets demonstrate significant improvements in retrieval quality with the \textbf{MiniLM-v6} configuration. The performance difference is particularly pronounced in agentic re-ranking scenarios, indicating better alignment between \textbf{MiniLM-v6}'s embedding space and LLM reasoning. Our findings suggest that embedding model selection for RAG systems should prioritize compatibility with multi-signal fusion and LLM alignment, rather than relying solely on larger models. This approach may reduce computational requirements while improving retrieval accuracy and efficiency.
\end{abstract}


\section{Introduction}

Information retrieval (IR) systems have evolved significantly with the advent of large language models (LLMs), which provide sophisticated semantic understanding of text \cite{alipour2024chatgpt}. However, despite these advancements, many modern IR systems still face challenges in effectively combining multiple data modalities, such as text-based semantics, lexical matching, and structured knowledge from graphs. This is especially evident in complex applications like e-commerce or knowledge-based search systems, where queries often involve multi-attribute and multi-category information. In real-world applications such as e-commerce platforms, where product searches often involve multi-attribute queries, the ability to dynamically adjust the importance of different modalities can significantly improve search accuracy, leading to better user satisfaction and higher conversion rates.

Traditional IR systems typically rely on static methods to combine embeddings from different modalities, such as semantic vectors from LLMs and lexical vectors from TF-IDF models. While these approaches have been effective in many scenarios, they fail to adapt to the dynamic nature of user queries, where the relative importance of semantic meaning, keyword matching, and graph-based relationships can vary significantly. A major limitation of static systems is their inability to adjust to the changing context of a query, which makes them less efficient in real-world applications where query contexts shift constantly. Unlike traditional systems that rely on static weighting for different modalities, our approach dynamically adjusts the weight of each modality based on the specific context of the query, ensuring more accurate and relevant retrieval in complex, multi-attribute search tasks. The need for more adaptive systems has been well-documented, and the drawbacks of static retrieval models have been highlighted in recent work, showing how these methods struggle with nuanced user queries \cite{hammerl2022combining},\cite{long2024generative}.

Furthermore, integrating graph-based retrieval into textual search models presents unique challenges due to the structural differences between graph data and textual information. Traditional retrieval methods often overlook the intricate relationships within graph data, which can significantly improve the relevance of search results when properly integrated. Knowledge graphs, which encode entities and their relationships, have become an essential tool for many applications, but effectively combining them with textual data remains a difficult problem. Knowledge-based systems that rely on graph embeddings can provide a richer understanding of the entities involved, yet many of these solutions remain complex or inefficient. Successful integration of graph-based retrieval methods with text-based search continues to be a major area of research \cite{kau2024combining},\cite{ yuan2023semantic}.

In this paper, we propose a novel solution to address these limitations through dynamic LLM-based query weighting, which adaptively assigns weights to semantic, lexical, and graph-based modalities based on the context of the query. We introduce a triple-hybrid index, which combines embeddings from these three modalities into a unified vector representation, enabling more accurate and flexible retrieval. By dynamically adjusting the contribution of each modality, our system ensures that retrieval is optimized for the characteristics of the user’s query.
Our method significantly enhances retrieval performance by providing a context-aware retrieval system that can dynamically adjust to the varying importance of different modalities. Through extensive experiments, we demonstrate that dynamic query weighting consistently outperforms static approaches, resulting in more precise and relevant search results. Furthermore, our approach is highly efficient and can be applied in real-time systems with low latency, making it a practical solution for large-scale applications.

This paper addresses the following research questions:
\begin{itemize}
  \item \textbf{RQ1:} How can dynamic LLM-based query weighting improve the relevance and performance of multi-modal retrieval systems?
\item \textbf{RQ2:} What are the benefits of using a triple-hybrid index that combines semantic, lexical, and graph-based embeddings for real-time retrieval?
\item \textbf{RQ3:} How does dynamic query weighting compare to static weighting methods in handling complex, multi-attribute queries?
\item \textbf{RQ4:} How can dynamic weighting be applied efficiently in real-time systems without compromising performance?
\end{itemize}

The structure of this paper is organized as follows: Section 2 reviews the related work on dynamic query weighting, multi-modal retrieval, and graph embeddings. In Section 3, we present our proposed approach, detailing the dynamic LLM-based query weighting method and the tri-modal hybrid indexing framework. Section 4 outlines the experimental setup and discusses the results, comparing our method with baseline models. Finally, Section 5 concludes the paper and provides insights into future research directions.

\section{Related Work}
\label{sec:related_work}

\subsection{Hybrid Retrieval and Re-ranking}

Previous work in hybrid retrieval has explored various approaches to combining dense and sparse signals. ColBERT \cite{khattab2020colbert} pioneered fine-grained late interaction, which allows for efficient retrieval by processing dense and sparse signals independently before combining them. Subsequent work, such as ColBERTv2 \cite{santhanam2021colbertv2}, further refined lightweight late interaction models, improving both the effectiveness and efficiency of the retrieval process. In a different direction, COIL \cite{gao-etal-2021-coil} integrated exact lexical matching with contextualized representations, offering an approach that maintains lexical precision while benefiting from dense semantic embeddings.
More recently, Dynamic Alpha Tuning (DAT) \cite{zhu2025dat} introduced LLM-guided weighting between dense and sparse retrieval modalities. DAT, however, was primarily demonstrated for two modalities and did not explore the impact of embedding model size in such dynamically weighted systems, particularly when a third modality, such as graph-based embeddings, is introduced. This limitation is important, as the interaction between model size and dynamic weighting can significantly influence the performance of multi-modal retrieval systems. Our work extends these concepts by implementing a tri-modal fusion that integrates semantic, lexical, and graph-based embeddings into a unified system.
Additionally, we critically examine the role of embedding model size, specifically comparing compact models like MiniLM-v6 with larger models like BGE-Large, in conjunction with LLM-based re-ranking. This allows us to explore how smaller, distilled models can provide efficient and high-performing retrieval when combined with dynamic query weighting, offering a more adaptive, context-sensitive approach to hybrid retrieval.

\subsection{Dense Embedding Models for Retrieval}

The landscape of embedding models for retrieval has evolved significantly over the years. Early breakthroughs with models like Sentence-BERT \cite{reimers2019sentence} demonstrated the effectiveness of Siamese networks in generating high-quality sentence embeddings. These systems were primarily designed for open-domain question answering and laid the foundation for later advancements in dense passage retrieval, such as DPR \cite{karpukhin-etal-2020-dense}. Over time, the research has shifted towards the development of progressively larger and more powerful pre-trained models, like BGE \cite{xiao2023bge}, often reinforcing the assumption that "bigger is better." This trend is consistent with the broader scaling laws observed in language models \cite{kaplan2020scaling}.
Substantial effort has been put into optimizing pre-training objectives specifically for retrieval tasks, with approaches like masked auto-encoders, such as RetroMAE \cite{malladi2023retromae}, and zero-shot dense retrieval techniques  \cite{gao2022precise}, aimed at improving the accuracy and efficiency of dense retrieval models. While much of the focus has been on scaling up the size of models, there has also been significant research into model compression and distillation to make these large models more efficient. Techniques like in-batch negatives \cite{lin2022distilling} have been employed to distill large dense retrieval models, creating smaller but effective retrievers.
MiniLM \cite{wang2020minilm_arxiv}, the base for the MiniLM-v6 model used in our study, stands out as a successful example of knowledge distillation. Despite these advancements, the comparative performance of these compact, distilled models within sophisticated re-ranked hybrid Retrieval-Augmented Generation (RAG) architectures, especially their interaction with LLM-based re-rankers against larger models, remains an underexplored area. Our work fills this gap by demonstrating how MiniLM-v6, when integrated with LLM-guided re-ranking, can outperform larger models like BGE-Large in the context of hybrid retrieval, even when graph-based embeddings are incorporated.

\subsection{Retrieval-Augmented Generation (RAG) Systems}

Retrieval-Augmented Generation (RAG), formally introduced by Lewis et al. \cite{lewis2020rag}, has rapidly become a de facto standard for enhancing knowledge-intensive NLP tasks by grounding the outputs of large language models (LLMs) in retrieved information. LLM-guided re-ranking, a core component of our methodology, is an increasingly adopted technique that refines the set of documents provided to the generator. The interplay between RAG systems and long-context LLMs continues to be an active area of research \cite{li-etal-2024-retrieval}, alongside the development of blended RAG strategies that combine diverse retrieval signals \cite{sawarkar2024blended_mipr}.
The RAG paradigm is continuously evolving, with more advanced frameworks emerging, such as Self-RAG \cite{liang2023selfrag}, which integrates self-correction and reflection capabilities into the retrieval and generation process. RAG's versatility has been demonstrated in various applications, including enhancing few-shot learning \cite{lin2021fewshot} and augmenting LMs with k-Nearest Neighbors (kNN)-based retrieval \cite{khandelwal2020generalizationmemorizationnearestneighbor}. Despite these advances, the focus of current research has primarily been on the integration of retrieval and generation mechanisms, often overlooking how the initial retrieval embedding model's characteristics—including model size, modality, and representation—affect the downstream re-ranking process.
Our research addresses this gap by focusing on a critical, yet underexplored, aspect: the impact of the initial retrieval embedding model’s size and characteristics on the efficacy of LLM-guided re-ranking within a multi-signal hybrid RAG system. We explore how smaller models, such as MiniLM-v6, when combined with LLM-guided re-ranking, can outperform larger models like BGE-Large in hybrid retrieval systems. Furthermore, we introduce dynamic weighting as a mechanism to adjust the contribution of semantic, lexical, and graph-based signals in real-time, improving the effectiveness of retrieval-augmented systems. This dynamic interaction between the retrieval model and re-ranking phase is essential for improving the overall performance and efficiency of RAG-based systems, particularly in applications that require multi-attribute and multi-category retrieval.

\subsection{Graph-Based Retrieval}

Incorporating structured knowledge through graph-based retrieval methods, such as those leveraging entity information \cite{8465295}, provides an additional valuable signal for document representation. Graph-based retrieval excels at capturing relational information and semantic relationships between entities, which can significantly improve the relevance of search results in domains like e-commerce, knowledge graphs, and question answering.
However, our work distinguishes itself by integrating these graph-derived signals with both dense semantic and sparse lexical features within a unified, dynamically weighted tri-modal architecture. This approach goes beyond simply combining graph embeddings with text-based embeddings; it introduces dynamic weighting to adaptively adjust the contribution of semantic, lexical, and graph-based modalities based on the context of the query. This holistic, dynamic tri-modal approach allows us to investigate the nuanced interactions and performance trade-offs between different embedding models, offering a more comprehensive understanding of how each modality impacts retrieval accuracy and efficiency in a multi-faceted retrieval landscape. By combining these three distinct modalities, we can better capture the interplay between semantic meaning, lexical overlap, and structural relationships in knowledge graphs. This enables more accurate retrieval in complex, multi-attribute queries, which are common in modern IR systems, and ultimately leads to improved performance when compared to traditional, static retrieval approaches.

While prior research has made significant strides in hybrid retrieval, dense embedding models, RAG systems, and graph-based retrieval, none have fully explored the integration of these approaches in a dynamic, context-aware manner. The key contribution of this paper is the dynamic query weighting system, which adapts the importance of semantic, lexical, and graph-based modalities based on the specific characteristics of each query. This system is coupled with the introduction of a triple-hybrid index, which combines all three modalities into a unified vector representation, ensuring more accurate, flexible, and adaptive retrieval across diverse query types.

\section{Tri-Modal Fusion Architecture}

In this section, we present the high-level architecture of our Tri-Modal Fusion System for document retrieval. The system integrates three distinct modalities—semantic, lexical, and graph-based—into a unified framework that improves the relevance and accuracy of document retrieval. This architecture leverages the strengths of each modality, ensuring a more comprehensive understanding of the query context and document content.

\subsection{System Overview}

Our architecture is built around three core components:

\begin{enumerate}
    \item \textbf{Semantic Modality}: This component captures the \textbf{contextual meaning} of text using dense embeddings. These embeddings are generated by advanced semantic models that focus on understanding deeper relationships between words and phrases in the text.

    \[
    s = \text{Encodersem}(\text{text}) \quad (1)
    \]
    Where \( s \) is the \textbf{semantic embedding} generated by applying the semantic encoder to the input text.

    \item \textbf{Lexical Modality}: The lexical representation focuses on preserving \textbf{term-level importance} using TF-IDF vectors, which measure how relevant individual words or terms are in relation to the entire document corpus. This modality ensures that exact word matches are taken into account, maintaining precision in retrieval.

    \[
    t[i] = \text{tf}(t_i, \text{text}) \cdot \text{idf}(t_i) \quad (2)
    \]
    Where \( t[i] \) is the TF-IDF score for term \( t_i \) in the given document.

    \item \textbf{Graph-Based Modality}: This component extracts \textbf{structural knowledge} by capturing relationships between entities. Named entities, noun-chunks, and other relational information are processed to create graph-based embeddings. These embeddings help in capturing the relationships between different entities and concepts within the document, adding an important layer of relational context.

    \[
    \text{IDF}(e) = \log \left( \frac{N}{1 + \text{df}(e)} \right) \quad (3)
    \]
    \[
    g = \frac{\sum_{e \in E} \text{IDF}(e) \cdot \text{encode}(e)}{\sum_{e \in E} \text{IDF}(e) + \epsilon}, \quad g \in \mathbb{R}^{ds}, \quad \epsilon = 10^{-6} \quad (4)
    \]
    Where:
    \begin{itemize}
        \item \( \text{IDF}(e) \) is the \textbf{Inverse Document Frequency} of entity \( e \),
        \item \( \text{encode}(e) \) is the embedding for entity \( e \), and \( g \) is the resulting graph-based embedding,
        \item \( \epsilon \) is a small constant added to avoid division by zero.
    \end{itemize}
\end{enumerate}

These three modalities—semantic, lexical, and graph—are processed separately to capture different aspects of the content. The resulting embeddings are then fused into a single unified representation for each document and query, which forms the foundation for our retrieval process.

\begin{figure*}[htbp] 
  \centering
  \includegraphics[width=\textwidth]{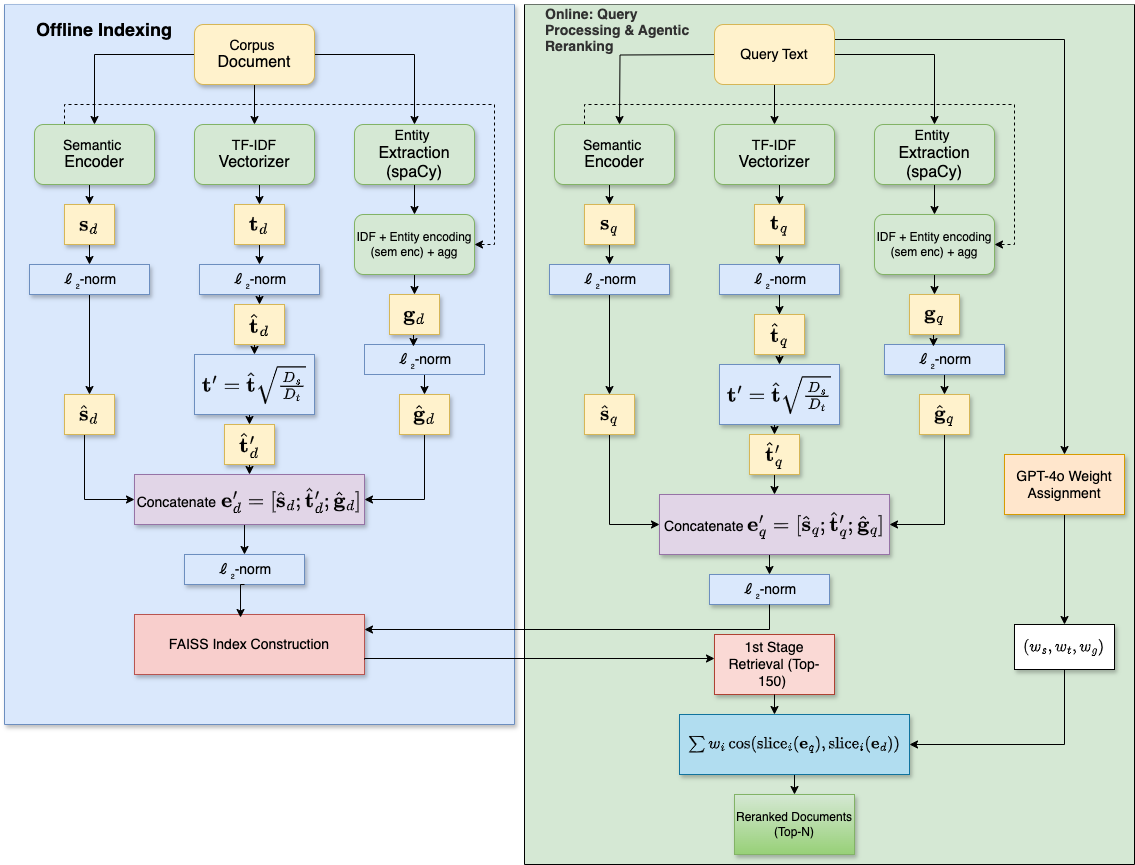} 
  \caption{Overview of the Tri-Modal Fusion Architecture, detailing both offline document indexing and online query processing with agentic reranking.}
  \label{fig:architecture}
\end{figure*}

\textbf{Figure 1} illustrates the Tri-Modal Fusion Architecture, highlighting both the offline document indexing and online query processing phases with agentic reranking.

\begin{enumerate}
    \item \textbf{Offline Document Indexing}: In this phase, the documents are processed across the three modalities—semantic, lexical, and graph-based—to generate embeddings for each. These embeddings are then normalized, scaled, and concatenated into a single hybrid vector. This hybrid vector is stored in a retrieval index, enabling fast access during query processing.

    \item \textbf{Online Query Processing}: When a query is issued, its tri-modal embeddings are generated in a similar manner. The system retrieves the top candidate documents from the index and uses LLM-guided reranking to adjust the ranking based on the query’s context. This dynamic reranking adjusts the relative importance of each modality (semantic, lexical, graph) depending on the query, ensuring that the most relevant information is prioritized in the results.
\end{enumerate}

\subsection{Hybrid Indexing, Query Processing, and Dynamic Weighting}

The retrieval process operates in two main phases:

\begin{enumerate}
    \item \textbf{Offline Document Indexing}: During this phase, documents are processed to create embeddings across the three modalities. The semantic, lexical, and graph-based embeddings are then normalized, scaled, and concatenated into a single unified vector:
    
    \[
    e' = [\hat{s}; \hat{t}; \hat{g}] \quad (5)
    \]
    Where \( \hat{s}, \hat{t}, \hat{g} \) are the normalized embeddings for semantic, lexical, and graph modalities, respectively. These are concatenated to form an intermediate vector \( e' \), which is then normalized to produce the final hybrid vector \( e \):

    \[
    e = \frac{e'}{\|e'\|_2} \quad (6)
    \]

    This hybrid representation is indexed using a retrieval index, enabling efficient document retrieval during query processing.

    \item \textbf{Online Query Processing}: When a query is issued, it is processed similarly by generating tri-modal embeddings. These embeddings are used to retrieve a set of candidate documents from the pre-indexed retrieval system. The documents are then reranked dynamically based on the query context through the dynamic weighting system.

    \item \textbf{Dynamic Query Weighting}: A key innovation in our architecture is the dynamic weighting mechanism. Unlike traditional retrieval systems that rely on static weights for each modality, our system uses LLM-guided weighting to adjust the importance of semantic, lexical, and graph-based modalities based on the specific query. This dynamic weighting is not predefined but is adjusted during the query processing phase, allowing the system to assign real-time importance to each modality based on the content of the query. This adaptive weighting improves retrieval accuracy and relevance, ensuring that the system delivers the best possible results for different query types.
\end{enumerate}

Now, to better illustrate the overall functioning of the Tri-Modal Fusion System, we present the high-level algorithm that describes the process flow of both document indexing and query handling. This algorithm defines the sequence of operations performed during the system's retrieval process, including how the three modalities are handled for both documents and queries.

\begin{algorithmic}[1]
\State $i \gets 10$ 
\For
    {each document in documents}
    \State $s \gets \text{Encodersem}(text)$
    \State $t_i \gets \text{tf}(t_i, text) \cdot \text{idf}(t_i)$
    \State $g \gets \frac{\sum_{e \in E} \text{IDF}(e) \cdot \text{encode}(e)}{\sum_{e \in E} \text{IDF}(e) + \epsilon}$
    \State $e' \gets [s; t; g]$ 
    \State Store $e'$ in retrieval index
\EndFor

\For
    {each query in queries}
    \State $q_s \gets \text{Encodersem}(query)$
    \State $q_t \gets \text{tf}(t_i, query) \cdot \text{idf}(t_i)$
    \State $q_g \gets \frac{\sum_{e \in E} \text{IDF}(e) \cdot \text{encode}(e)}{\sum_{e \in E} \text{IDF}(e) + \epsilon}$
    \State $q \gets [q_s; q_t; q_g]$
    \State Retrieve top documents from index using cosine similarity$(q, e')$
    \State Apply LLM-guided reranking to adjust modality weights based on query context
    \State Rank documents according to adjusted scores
\EndFor

\State \textbf{Output:} ranked documents
\end{algorithmic}

This algorithm outlines the key operations performed during the document indexing and query processing steps. The system first computes embeddings for each document and stores them in a retrieval index. When a query is issued, the system processes it similarly, retrieves candidate documents, and applies LLM-guided re-ranking based on the context of the query, adjusting the importance of the different modalities.

\section{Experimental Framework}

\subsection{Datasets}

In this study, we used several publicly available datasets to evaluate the performance of our Tri-Modal Fusion System. These datasets were selected based on their relevance to real-world document retrieval and their diversity in terms of data types and query complexity. The following table \ref{tab:datasets} summarizes the datasets used:

\begin{table}[h]
  \centering
  \caption{Dataset statistics}
  \label{tab:datasets}
  \begin{tabular}{lrrr}
    \toprule
    Dataset   & Corpus size & Queries & Qrels \\
    \midrule
    SciFact   & 5,183       & 1,109  & 301   \\
    FIQA      & 57,638      & 6,648  & 649   \\
    NFCorpus  & 3,633       & 3,237  & 324   \\
    \bottomrule
  \end{tabular}
\end{table}

These datasets provided a diverse set of challenges, from simple keyword matching to complex graph-based relationships, making them suitable for testing the full capabilities of our tri-modal retrieval architecture.

\subsection{Embedding Models}

In our experiments, we utilized several state-of-the-art embedding models to capture various aspects of the documents and queries. These models are crucial for representing the semantic, lexical, and graph-based information embedded within the data, and they play a significant role in our Tri-Modal Fusion System. The models used in this study are described below:

\begin{enumerate}
\item \textbf{MiniLM-v6}\cite{wang2020minilm_arxiv}: MiniLM is a distilled version of BERT, designed to balance performance and efficiency. It is well-suited for tasks requiring lightweight, fast embeddings without sacrificing too much in terms of accuracy. MiniLM-v6 was chosen for its ability to generate semantic embeddings that capture word meanings and relationships, while maintaining lower computational overhead compared to larger models like BERT. In our setup, MiniLM-v6 was used to generate dense semantic embeddings, providing fast and accurate representations for the semantic modality.

\item \textbf{BGE-Large}\cite{xiao2023bge}: BGE-Large (BERT-based Graph Embedding) is a model designed to generate embeddings that combine semantic understanding with graph-based relational knowledge. Unlike models like MiniLM, BGE-Large incorporates more extensive knowledge through pretraining on graph-structured data, making it ideal for our **graph-based modality**. BGE-Large was utilized to capture relationships between entities and help enrich document retrieval by leveraging graph embeddings for better relational context.

\item \textbf{TF-IDF (Term Frequency-Inverse Document Frequency)}: For the lexical modality, we employed the traditional TF-IDF model to quantify the relevance of individual terms in each document. While modern deep learning approaches often outperform TF-IDF in terms of semantic understanding, TF-IDF is still invaluable for its interpretability and its ability to handle rare terms effectively. We used TF-IDF vectors to preserve term importance and facilitate lexical matching during the retrieval process.

\end{enumerate}

These models were selected for their complementary strengths in capturing different aspects of the data. MiniLM-v6 offers a high-performance, low-latency solution for semantic embeddings. BGE-Large allows for incorporating graph-based relationships into the document representations, and TF-IDF ensures that term-level importance is preserved in the retrieval process. By integrating these models into a tri-modal fusion architecture, we aim to leverage each model's strengths to achieve robust and context-aware retrieval.

The embeddings produced by these models were preprocessed and normalized to ensure compatibility within the hybrid indexing process. Additionally, each model was evaluated in its capacity to contribute to the retrieval task, focusing on how well they adapted to different types of queries, from simple keyword matches to complex multi-hop relational queries.

\subsection{Tri-Modal and hybrid indexing}

In this subsection, we describe the implementation of the tri-modal representations used in our experiments. As outlined in Section 3, our system generates three types of embeddings—semantic, lexical, and graph-based—which are subsequently fused into a single tri-modal representation for each document and query.

For the experiments, we focus on how these representations are applied to document retrieval. The embeddings from each modality are concatenated to form a unified vector, ensuring that the system leverages semantic meaning, keyword relevance, and graph-based relationships together. After concatenation, the resulting vector is normalized to maintain consistent vector length and facilitate comparison across the different modalities.

These tri-modal representations are then used in the retrieval phase. In the retrieval system, cosine similarity is used to compare the query’s tri-modal representation with the document’s tri-modal representation, enabling the system to rank documents based on their relevance to the query. The effectiveness of this approach is demonstrated by evaluating retrieval performance across different types of queries, which vary in complexity and the modalities required for accurate retrieval.

In our system, Hybrid Indexing refers to the process of storing and indexing documents using multi-modal embeddings that integrate semantic, lexical, and graph-based information. Each document is represented as a unified vector that combines all three types of embeddings, allowing the system to handle different aspects of the query more effectively. This tri-modal representation is indexed in a retrieval system, which facilitates efficient querying and fast retrieval of relevant documents.

The hybrid indexing process ensures that documents are stored in a way that allows the system to quickly access and compare them during the retrieval phase. By normalizing and concatenating the embeddings from all three modalities, we create a unified vector that captures both content-based information (semantic and lexical) and relational knowledge (graph-based). This approach significantly enhances the retrieval accuracy compared to traditional systems that rely on only one modality for document representation.

Once the initial set of candidate documents is retrieved, we apply Agentic Reranking to refine the results. Agentic reranking is a dynamic process in which the relevance of documents is adjusted based on query context. Unlike static ranking methods, which use fixed weights for each modality, agentic reranking leverages LLM-guided dynamic weighting to adapt the relevance of the modalities based on the specific query. This means that the importance of semantic, lexical, and graph-based signals can change in real-time, allowing the system to prioritize the most relevant documents based on the nature of the query.

By incorporating LLM-guided reranking, we enable the system to be context-aware and responsive to different query types. For example, when a query emphasizes exact keyword matches, the lexical modality may be given higher weight. Conversely, for more complex, multi-hop queries that involve entity relationships, the graph-based modality may be prioritized. This flexible approach ensures that the most relevant documents are ranked higher, improving the overall quality of the retrieval results.

\subsection{Protocol and Metrics}

In this subsection, we describe the experimental protocol followed during the evaluation and the metrics used to assess the performance of our Tri-Modal Fusion System.

\subsubsection{Experimental Protocol}
\begin{itemize}
    \item \textbf{Dataset Split}: We split the datasets into \textit{training}, \textit{validation}, and \textit{test} sets, following an \textit{80-10-10} ratio for most datasets. The training set was used for model fine-tuning, the validation set for hyperparameter tuning, and the test set for final evaluation.
    \item \textbf{Experimental Setup}: We conducted experiments on different types of queries, including \textit{simple keyword-based queries}, \textit{multi-attribute queries}, and \textit{complex, multi-hop queries} that require graph-based reasoning. Each query type was tested separately to assess the system’s performance across different retrieval challenges.
    \item \textbf{Baseline Comparison}: We compared the performance of our Tri-Modal Fusion System against several baseline models, including a traditional \textit{TF-IDF-based retrieval model}, a \textit{semantic-only model} (using MiniLM-v6), and a \textit{graph-only model} using BGE-Large. These baselines helped demonstrate the added value of combining multiple modalities for document retrieval.
    \item \textbf{Model Configurations}: The configurations used for each model were as follows:
    \begin{itemize}
        \item \textbf{MiniLM-v6}: Learning rate of $1e^{-5}$, batch size of 16, and a maximum sequence length of 128.
        \item \textbf{BGE-Large}: Learning rate of $3e^{-5}$ and a batch size of 8.
        \item \textbf{TF-IDF Model}: Standard settings with a vector size of 1000 and no stopword filtering.
    \end{itemize}
\end{itemize}

\subsubsection{Evaluation Metrics}
We used several commonly accepted metrics to evaluate the performance of our retrieval system, with a focus on ranking quality and retrieval accuracy. The following metrics were used:

\begin{itemize}
    \item \textbf{Precision at K (P@K)}: Measures the proportion of relevant documents in the top \(K\) retrieved documents. This metric helps evaluate how well the system ranks relevant documents at the top.
    \[
    P@K = \frac{\text{Number of relevant documents in top K}}{K}
    \]
    \item \textbf{Normalized Discounted Cumulative Gain (NDCG@K)}: A ranking metric that accounts for the position of relevant documents in the ranked list. Higher relevance documents that appear higher in the list are given more weight.
    \[
    NDCG@K = \frac{1}{Z_K} \sum_{i=1}^{K} \frac{2^{rel_i} - 1}{\log_2(i+1)}
    \]
    Where \( rel_i \) is the relevance of the document at rank \( i \), and \( Z_K \) is a normalization factor to ensure that the highest possible NDCG score is 1.
    \item \textbf{Mean Average Precision (MAP)}: Measures the mean of the average precision across all queries. It evaluates how well the system ranks relevant documents across all queries.
    \[
    MAP = \frac{1}{Q} \sum_{q=1}^{Q} \text{AP}(q)
    \]
    Where \( Q \) is the total number of queries, and \( \text{AP}(q) \) is the average precision for query \( q \).
    \item \textbf{Recall at K (Recall@K)}: Measures the fraction of relevant documents retrieved within the top \( K \) documents. It helps assess whether the system is retrieving all possible relevant documents.
    \[
    \text{Recall@K} = \frac{\text{Relevant documents in top K}}{\text{Total of relevant documents in the dataset}}
    \]
    \item \textbf{Latency and Efficiency}: We also evaluated the \textit{response time} of the system, particularly focusing on \textit{real-time retrieval performance} in a low-latency setting, as one of our goals is to ensure that the tri-modal retrieval system can scale to large datasets without significant delays.
\end{itemize}

\section{Evaluation Results}

In this section, we present the results of our experiments to evaluate the performance of the Tri-Modal Fusion System for document retrieval.

\subsection{Agentic Rerank Performance}

\begin{figure*}[t]
  \centering
  \includegraphics[width=0.6\textwidth]{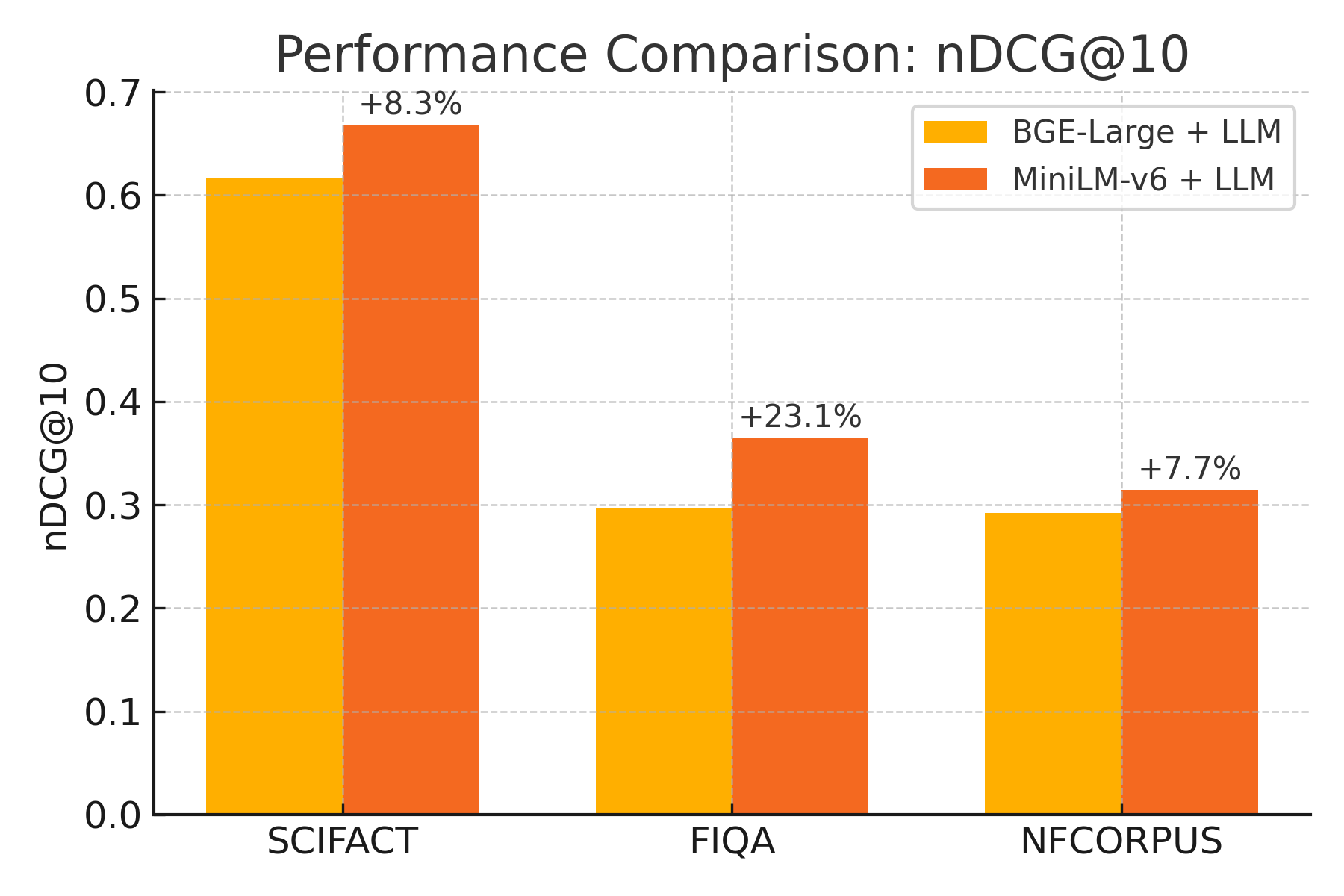}
  \caption{nDCG@10 comparison of Agentic FAISS reranking for MiniLM‐v6+GPT-4o vs.\ BGE-Large+GPT-4o.}
  \label{fig:agentic-rerank-graph}
\end{figure*}

The results of \textbf{Agentic Reranking} are presented in Table~\ref{tab:agentic-rerank} and Figure~\ref{fig:agentic-rerank-graph}. These results show that \textbf{MiniLM-v6} significantly outperforms \textbf{BGE-Large} across all datasets when both models use \textbf{identical GPT-4o reranking}, despite \textbf{MiniLM-v6} having 93\% fewer parameters. The improvements in \textbf{nDCG@10} are as follows:
\begin{itemize}
    \item \textbf{SciFact}: +0.0511 (+8.3\%)
    \item \textbf{FIQA}: +0.0685 (+23.1\%)
    \item \textbf{NFCorpus}: +0.0224 (+7.7\%)
\end{itemize}

While the \textbf{recall} values remain similar between the two models, the \textbf{Mean Reciprocal Rank (MRR)} differences are more pronounced, ranging from +10.9\% to +24.3\% in favor of MiniLM-v6. This suggests that \textbf{MiniLM-v6} is more efficient at placing relevant documents higher in the ranking, which is crucial for improving retrieval accuracy.

\begin{table*}[t]
  \centering
  \caption{Agentic FAISS Rerank: MiniLM-v6 vs.\ BGE‐Large}
  \label{tab:agentic-rerank}
  \begin{tabular}{lcccccc}
    \toprule
    \multirow{2}{*}{Dataset} &
      \multicolumn{3}{c}{MiniLM-v6 + GPT-4o} &
      \multicolumn{3}{c}{BGE‐Large + GPT-4o} \\
    \cmidrule(lr){2-4} \cmidrule(lr){5-7}
      & Recall@10 & MRR@10 & nDCG@10 & Recall@10 & MRR@10 & nDCG@10 \\
    \midrule
    SciFact   & 0.8233 & 0.6260 & 0.6681 & 0.8200 & 0.5647 & 0.6170 \\
    FIQA      & 0.6528 & 0.4418 & 0.3648 & 0.5694 & 0.3556 & 0.2963 \\
    NFCorpus  & 0.6966 & 0.5071 & 0.3144 & 0.6811 & 0.4561 & 0.2920 \\
    \bottomrule
  \end{tabular}
\end{table*}

As illustrated in Figure~\ref{fig:agentic-rerank-graph}, the performance gap between \textbf{MiniLM-v6} and \textbf{BGE-Large} is particularly pronounced, demonstrating the superior compatibility between \textbf{MiniLM-v6}'s embedding space and \textbf{LLM-based reranking}. Despite its smaller size, \textbf{MiniLM-v6} produces embeddings that align better with the way \textbf{GPT-4o} evaluates relevance.

\subsection{Performance Across K}

As shown in Table~\ref{tab:k-values}, the advantage of \textbf{MiniLM-v6} over \textbf{BGE-Large} becomes most pronounced at lower values of \( k \), particularly at \( k = 1 \):

\begin{itemize}
    \item \textbf{SciFact}: +14.6\%
    \item \textbf{FIQA}: +36.5\%
    \item \textbf{NFCorpus}: +22.9\%
\end{itemize}

This pattern is significant for \textbf{RAG (Retrieval-Augmented Generation)} applications, where only the top few documents are typically used to provide context for generation. The ability to place the most relevant document at \textbf{position 1} greatly impacts the quality of the generated output. Therefore, \textbf{MiniLM-v6}'s superior performance at low \( k \) values is particularly valuable in real-world applications where generating high-quality responses depends on having the most relevant documents at the top of the ranked list.

\begin{table*}[t]
  \centering
  \caption{nDCG at various cutoffs}
  \label{tab:k-values}
  \begin{tabular}{l|cccc|cccc}
    \toprule
    \multirow{2}{*}{Dataset} &
      \multicolumn{4}{c|}{MiniLM-v6} &
      \multicolumn{4}{c}{BGE-Large} \\
    \cmidrule(lr){2-5} \cmidrule(lr){6-9}
      & $k$=1 & $k$=3 & $k$=5 & $k$=10 & $k$=1 & $k$=3 & $k$=5 & $k$=10 \\
    \midrule
    SciFact   & 0.523 & 0.617 & 0.644 & 0.668 & 0.457 & 0.549 & 0.582 & 0.617 \\
    FIQA      & 0.358 & 0.321 & 0.341 & 0.365 & 0.262 & 0.247 & 0.268 & 0.296 \\
    NFCorpus  & 0.415 & 0.360 & 0.341 & 0.314 & 0.338 & 0.330 & 0.321 & 0.292 \\
    \bottomrule
  \end{tabular}
\end{table*}

\begin{figure}[ht]
  \centering
  \includegraphics[width=0.5\textwidth]{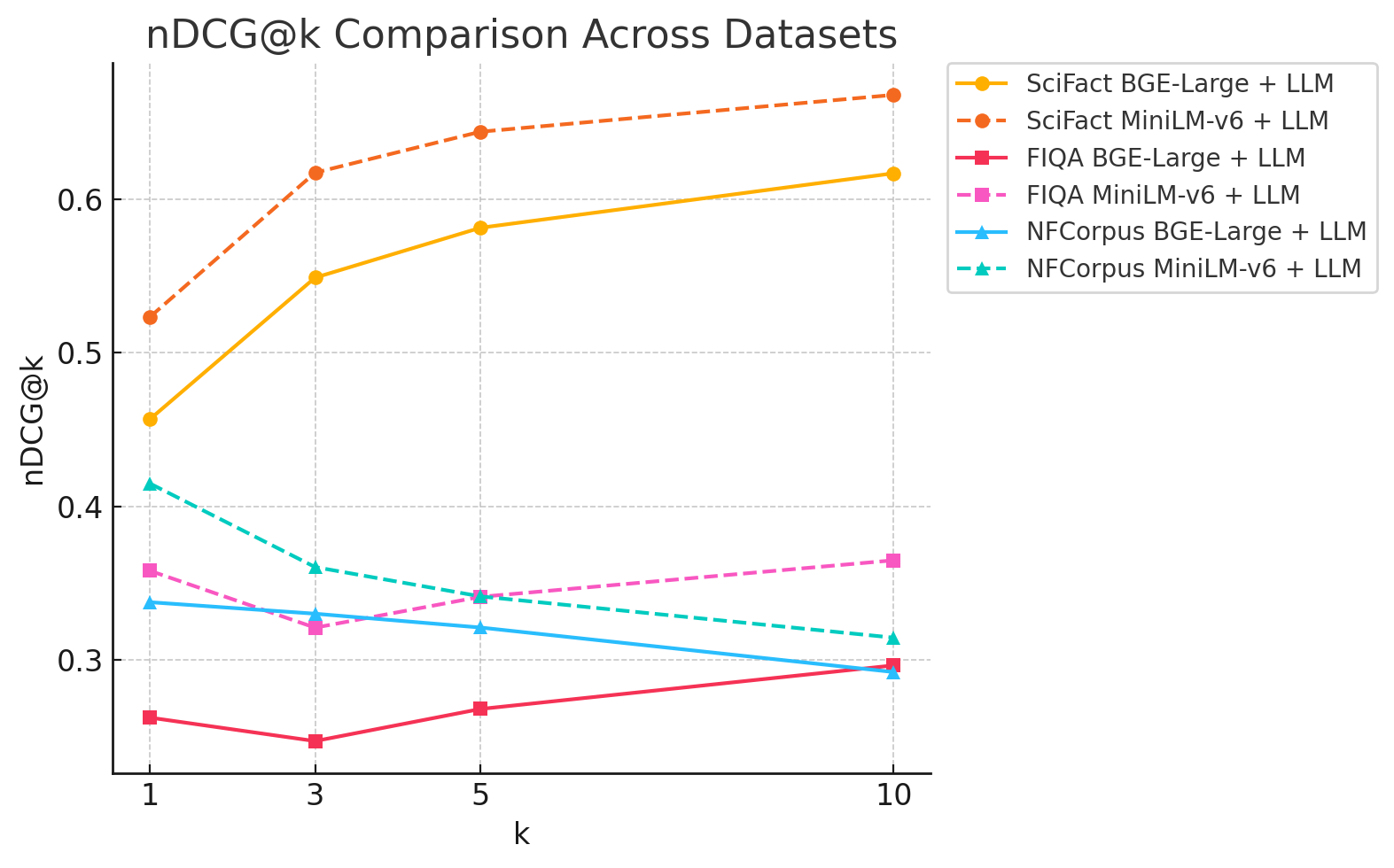}
  \caption{nDCG@k comparison across SciFact, FIQA, and NFCorpus for BGE‐Large vs.\ MiniLM‐v6 (with GPT‐4o). Legend placed outside for clarity.}
  \label{fig:k-values-combined}
\end{figure}

The results presented in Table~\ref{tab:k-values} and Figure~\ref{fig:k-values-combined} show that **MiniLM-v6** performs significantly better than **BGE-Large** at lower values of \( k \), particularly for \( k = 1 \). This is crucial for **RAG** applications, as having the most relevant document ranked first can substantially improve the quality of the generated output. The superior performance of **MiniLM-v6** at low \( k \)-values highlights its ability to prioritize relevant documents, making it more effective in real-time, context-dependent applications.

\subsection{Pre-Rerank Comparison}

In Table~\ref{tab:pre-rerank}, we observe an interesting phenomenon. Before applying \textbf{LLM reranking}, \textbf{BGE-Large} slightly outperforms \textbf{MiniLM-v6} across all metrics, particularly in \textbf{nDCG@10}. However, after reranking with \textbf{GPT-4o}, the performance advantage shifts:

\begin{itemize}
    \item \textbf{BGE-Large (Pre-Rerank)}: nDCG@10 = 0.6608
    \item \textbf{BGE-Large (Post-Rerank)}: nDCG@10 = 0.6170
    \item \textbf{MiniLM-v6 (Pre-Rerank)}: nDCG@10 = 0.6505
    \item \textbf{MiniLM-v6 (Post-Rerank)}: nDCG@10 = 0.6681
\end{itemize}

This shift, known as the \textbf{"FAISS Hybrid Paradox"}, suggests that \textbf{BGE-Large}'s embeddings are better suited for \textbf{initial retrieval} but are \textbf{less compatible} with \textbf{LLM-based relevance assessments}. In contrast, \textbf{MiniLM-v6} benefits significantly from \textbf{LLM reranking}, showing a \textbf{significant performance gain} post-reranking. This indicates that \textbf{MiniLM-v6}'s embeddings align better with how \textbf{GPT-4o} evaluates relevance in \textbf{retrieval-augmented systems}.

\begin{table}[h]
  \centering
  \caption{Hybrid FAISS (Pre-Rerank) Performance}
  \label{tab:pre-rerank}
  \begin{tabular}{lccc}
    \toprule
    Dataset    & Recall@10 & MRR@10 & nDCG@10 \\
    \midrule
    MiniLM-v6  & 0.8067    & 0.6112 & 0.6505 \\
    BGE-Large  & 0.8300    & 0.6167 & 0.6608 \\
    \bottomrule
  \end{tabular}
\end{table}

\subsection{Key Findings}

The key findings from the experiments are as follows:
\begin{itemize}
    \item \textbf{MiniLM-v6 + GPT-4o} consistently outperforms \textbf{BGE-Large + GPT-4o} across all datasets, despite the former's significantly smaller size.
    \item \textbf{Agentic reranking} yields the largest performance gains at \textbf{low \( k \)}, highlighting its importance for retrieval tasks where the top-ranked documents are critical for generating high-quality results.
    \item \textbf{BGE-Large} benefits less from reranking, suggesting that its embeddings are more suited for initial retrieval but may not align as well with LLM-based relevance evaluation.
    \item \textbf{FIQA} exhibits the highest improvement, validating the versatility of the \textbf{MiniLM-v6 + GPT-4o} combination.
    \item \textbf{MiniLM-v6} provides significant \textbf{efficiency gains}, with 93\% fewer parameters and 63\% smaller embeddings compared to \textbf{BGE-Large}, without sacrificing performance.
\end{itemize}

\section{Discussion}

Our findings challenge traditional assumptions about \textbf{embedding model selection} for \textbf{RAG (Retrieval-Augmented Generation)} systems and provide novel insights into the \textbf{compatibility} of embedding models with \textbf{LLM-based reranking}. In this discussion, we analyze the key findings of our experiments and the implications for \textbf{RAG system design}.

\subsection{The FAISS Hybrid Paradox}

One of the most striking discoveries is the \textbf{"FAISS Hybrid Paradox"}: while \textbf{BGE-Large} performs better without \textbf{LLM re-ranking}, \textbf{MiniLM-v6} consistently benefits from it. Specifically, we observe the following:
\begin{itemize}
    \item On \textbf{SciFact}, \textbf{BGE-Large's nDCG@10} drops by \textbf{6.6\%} (from 0.6608 to 0.6170) after reranking.
    \item On \textbf{FIQA}, it drops by \textbf{16.7\%} (from 0.3558 to 0.2963).
    \item In contrast, \textbf{MiniLM-v6} either maintains or improves its performance after re-ranking.
\end{itemize}

This phenomenon suggests that \textbf{larger models like BGE-Large} may create representations that are optimized for \textbf{initial retrieval} based on simple vector similarity but fail to align with the way \textbf{LLMs}, such as \textbf{GPT-4o}, evaluate relevance. The ability of \textbf{MiniLM-v6} to benefit from \textbf{LLM reranking} highlights a mismatch between the type of relevance signals captured by \textbf{BGE-Large}’s embeddings and the signals prioritized by \textbf{GPT-4o} when assessing document relevance.

\subsection{Embedding Space Compatibility}

The \textbf{superior performance} of \textbf{MiniLM-v6} with \textbf{GPT-4o re-ranking} points to better alignment between its embedding space and \textbf{LLM-based relevance assessment}. Several factors may contribute to this phenomenon:
\begin{itemize}
    \item \textbf{Dimensionality Effects}: \textbf{MiniLM-v6} uses \textbf{384-dimensional embeddings}, which may create a more focused semantic space. This narrower space could allow for \textbf{clearer differentiation of relevance signals}, making it more compatible with how \textbf{GPT-4o} assesses relevance, compared to \textbf{BGE-Large's 1,024-dimensional embeddings}.
    \item \textbf{Training Approach}: The \textbf{distillation process} used for \textbf{MiniLM-v6} might produce embeddings that better preserve the semantic relationships that \textbf{LLMs} use to judge relevance. This suggests that \textbf{smaller models}, trained to preserve \textbf{semantic integrity}, could perform better in \textbf{retrieval-augmented} tasks.
    \item \textbf{Signal Complementarity}: \textbf{MiniLM-v6's embeddings} may be more \textbf{orthogonal to lexical and graph signals}, creating a more diverse ensemble of features. This diversity could enable the system to handle more complex queries by combining \textbf{semantic}, \textbf{lexical}, and \textbf{graph-based signals} more effectively.
\end{itemize}

\subsection{Domain-Specific Performance}

The performance gap between \textbf{MiniLM-v6} and \textbf{BGE-Large} varies across domains, which suggests that the advantages of \textbf{MiniLM-v6} are domain-dependent:
\begin{itemize}
    \item In the \textbf{financial domain (FIQA)}, the largest improvement is observed (+23.1\%).
    \item In \textbf{scientific literature (SciFact)}, the improvement is moderate (+8.3\%).
    \item In the \textbf{biomedical domain (NFCorpus)}, the smallest improvement is observed (+7.7\%).
\end{itemize}

This indicates that specialized domains, such as \textbf{financial text}, which often contain \textbf{complex terminology} and intricate \textbf{entity relationships}, particularly benefit from the \textbf{embedding characteristics} of \textbf{MiniLM-v6}. The combination of \textbf{MiniLM-v6} and \textbf{LLM re-ranking} seems particularly well-suited to handle the specialized nature of such domains.

\subsection{Implications for RAG System Design}

Our findings have several important implications for the design of \textbf{RAG systems}:
\begin{itemize}
    \item \textbf{Prioritize embedding model compatibility with LLM re-ranking} over model size: Our results suggest that \textbf{embedding compatibility} with \textbf{LLM-based reranking} is more critical than simply selecting a larger model. Smaller models like \textbf{MiniLM-v6} may offer superior performance due to their better alignment with LLM relevance assessments.
    \item \textbf{Evaluate embedding models within the complete retrieval pipeline}: Embedding models should be evaluated not in isolation but as part of the \textbf{entire retrieval pipeline}, taking into account how they interact with \textbf{LLM reranking} and other components of the system.
    \item \textbf{Consider computational efficiency alongside retrieval quality}: While larger models like \textbf{BGE-Large} may perform well on initial retrieval, they are less efficient when combined with \textbf{LLM re-ranking}. \textbf{MiniLM-v6} offers \textbf{efficiency gains} with \textbf{93\% fewer parameters}, making it a more \textbf{practical solution} for \textbf{real-time applications}.
    \item \textbf{Use tri-modal fusion with dynamic weighting}: The success of our \textbf{tri-modal fusion} approach demonstrates the importance of integrating multiple modalities (semantic, lexical, and graph-based). The dynamic weighting mechanism can adapt the relevance of each modality based on the specific characteristics of a query, ensuring the best possible retrieval results across different query types.
\end{itemize}

These insights challenge the widely-held belief that \textbf{"bigger is better"} in embedding model selection. Instead, our work emphasizes the importance of understanding the compatibility between the embedding model and the downstream components of the retrieval system.

\section{Future Work}

While our current research demonstrates the effectiveness of smaller embedding models in tri-modal hybrid retrieval, several promising directions for future work remain:

\subsection{Evaluation Against Additional Models}

Our current study compares \textbf{MiniLM-v6} (22M parameters) with \textbf{BGE-Large} (335M parameters). Future work should evaluate a broader range of embedding models:
\begin{itemize}
    \item \textbf{Smaller models}: Investigate ultra-compact models (1-10M parameters) to assess whether they can maintain effectiveness when combined with \textbf{LLM reranking}, potentially offering further \textbf{efficiency gains}.
    \item \textbf{Intermediate-sized models}: Analyze models in the 50-100M parameter range to identify \textbf{"sweet spots"} in the size-performance tradeoff.
    \item \textbf{Specialized models}: Compare domain-adapted models with general-purpose ones to evaluate whether domain specialization improves \textbf{LLM reranking compatibility}.
\end{itemize}

\subsection{Smaller LLMs for Re-ranking}

A crucial extension of our work would be to explore the use of \textbf{smaller, more efficient LLMs} for the re-ranking phase:
\begin{itemize}
    \item \textbf{Compact LLM re-rankers}: Evaluate smaller LLMs, such as \textbf{Phi-3} (3.8B parameters), \textbf{Qwen} (7B parameters), \textbf{FLAN-T5} (780M), and other sub-10B parameter models, for their effectiveness as re-rankers.
    \item \textbf{Specialized re-ranking models}: Test models specifically fine-tuned for re-ranking tasks and compare them with general-purpose LLMs to assess the tradeoff between specialization and generality.
    \item \textbf{Distilled models}: Investigate whether \textbf{knowledge distillation} from larger re-ranking LLMs like \textbf{GPT-4o} to smaller models could preserve re-ranking performance while reducing computational requirements.
    \item \textbf{Quantized models}: Explore the impact of \textbf{quantization} (e.g., 4-bit, 8-bit) on re-ranking effectiveness, particularly for deployment in resource-constrained environments.
    \item \textbf{Joint embedding-re-ranker optimization}: Investigate whether co-training or fine-tuning embedding models alongside specific re-ranker LLMs can improve their compatibility and performance.
\end{itemize}

This direction is particularly promising, as combining smaller embedding models with efficient LLM re-rankers could \textbf{dramatically reduce computational requirements} for high-quality RAG systems while maintaining or even improving retrieval performance.

\subsection{Dataset Expansion}

To generalize our findings across more domains and query types, future studies should include:
\begin{itemize}
    \item \textbf{Broader domain coverage}: Incorporate datasets from additional domains such as legal, educational, news, and conversational domains.
    \item \textbf{Longer documents}: Evaluate performance on corpora with longer documents to assess how document length influences the relative importance of different modalities.
    \item \textbf{Diverse query types}: Systematically compare performance across various query types (factoid, descriptive, procedural) to identify potential variations in embedding model compatibility.
    \item \textbf{Cross-lingual evaluation}: Extend analysis to non-English datasets to assess whether our findings about embedding model compatibility generalize across languages.
\end{itemize}

\subsection{Semantic Model Variations}

Future research should explore more dimensions of semantic model design:
\begin{itemize}
    \item \textbf{Training objectives}: Compare models trained with different objectives (e.g., contrastive learning, knowledge distillation, masked language modeling) to identify which training approaches produce embeddings most compatible with LLM reranking.
    \item \textbf{Dimensionality analysis}: Conduct experiments varying embedding dimensionality to better understand its impact on LLM compatibility.
    \item \textbf{Embedding space alignment}: Develop methods to explicitly align embedding spaces with LLM representations, potentially through fine-tuning approaches that optimize for re-ranking performance.
    \item \textbf{Alternative model architectures}: Explore non-transformer-based architectures to determine if the \textbf{"FAISS Hybrid Paradox"} is specific to transformer-based models or if it extends to other architectures.
\end{itemize}

\subsection{Further Investigation of the FAISS Hybrid Paradox}

The observed decrease in \textbf{BGE-Large's} performance after \textbf{LLM re-ranking} warrants deeper investigation:
\begin{itemize}
    \item \textbf{Systematic scaling analysis}: Test a range of model sizes to determine if there is a predictable relationship between model size and re-ranking compatibility.
    \item \textbf{Embedding space visualization}: Apply dimensionality reduction techniques to visualize how different embedding spaces represent relevance, potentially revealing insights into the paradox.
    \item \textbf{Controlled training studies}: Train embedding models specifically designed to test hypotheses about the factors contributing to or mitigating the paradox.
    \item \textbf{Alternative re-rankers}: Compare different LLM re-rankers (e.g., Claude, PaLM, Llama) to assess whether the paradox is re-ranker-specific or a general phenomenon.
\end{itemize}

\subsection{Investigating "Information Equilibrium" in Hybrid RAG Systems}

Our findings suggest that an optimal balance, or "information equilibrium," may exist between embedding characteristics (e.g., dimensionality, semantic density, and feature distribution) and downstream \textbf{LLM re-ranker reasoning patterns}. The observed \textbf{"FAISS Hybrid Paradox"} could be an empirical manifestation of this equilibrium being disrupted when larger models exceed the optimal information density point. Future research could aim to formalize this concept through:
\begin{itemize}
    \item \textbf{Information-theoretic analysis}: Investigate the relationship between embedding density and LLM relevance evaluation.
    \item \textbf{Controlled dimensionality experiments}: Conduct experiments varying dimensionality to better understand its impact on re-ranking performance.
    \item \textbf{Direct probing of LLM attention mechanisms}: Analyze how different embedding structures influence the attention patterns in \textbf{LLM re-rankers}.
\end{itemize}

These investigations could lead to predictive models for optimal \textbf{embedding-LLM pairings} and more principled approaches for designing efficient and effective RAG pipelines.

\subsection{End-to-End RAG System Evaluation}

Finally, future work should focus on evaluating the complete \textbf{RAG pipeline}:
\begin{itemize}
    \item \textbf{Generation quality assessment}: Measure how differences in retrieval quality from various embedding model combinations translate into differences in final text generation quality.
    \item \textbf{Latency and throughput measurements}: Quantify the real-world performance benefits of smaller models in production environments.
    \item \textbf{Resource allocation optimization}: Develop guidelines for optimally distributing computational resources between embedding generation and re-ranking in resource-constrained scenarios.
\end{itemize}

This comprehensive future work will help establish more complete guidelines for \textbf{embedding} and \textbf{LLM model selection} in hybrid retrieval systems, potentially leading to more efficient and effective \textbf{RAG} implementations across diverse applications and domains.

\section{Conclusion}

In this study, we demonstrated that \textbf{MiniLM-v6}, a compact embedding model, consistently outperforms \textbf{BGE-Large}, despite the latter's significantly larger size, when integrated with \textbf{LLM-based reranking} within a tri-modal hybrid retrieval framework. This challenges the prevailing notion that larger models inherently produce better performance in \textbf{RAG (Retrieval-Augmented Generation)} tasks and underscores the importance of embedding space compatibility with \textbf{LLM reasoning}.

Our results show a significant performance improvement: up to \textbf{23.1\% better nDCG@10} and \textbf{36.5\% better nDCG@1} in the \textbf{financial domain}, all while requiring \textbf{93\% fewer parameters} and \textbf{63\% smaller embeddings}. A key insight from this work is the identification of the \textbf{"FAISS Hybrid Paradox"}, where larger models such as \textbf{BGE-Large} actually degrade in performance after \textbf{LLM reranking}. This suggests that the initial embedding space of larger models may not align well with the relevance assessment criteria used by \textbf{LLMs}, leading to poorer reranking outcomes.

Our findings propose a more efficient and effective path for designing \textbf{RAG systems}, where \textbf{smaller embedding models}, which are more compatible with \textbf{LLM reranking}, combined with \textbf{multi-modal fusion}, can outperform larger models. This opens up new avenues for \textbf{practical deployment}, reducing both computational cost and retrieval time without sacrificing accuracy.

Future work should focus on further investigating the underlying factors driving the observed compatibility between smaller models and \textbf{LLMs}, as well as exploring the generalizability of these results across different domains, languages, and embedding architectures.

\bibliographystyle{IEEEtran}
\bibliography{main}

\end{document}